\documentclass[12pt,thmsa]{article}
\usepackage{sw20lart}
\usepackage{graphicx}
\usepackage{epsfig}
\usepackage{setspace}

\input tcilatex
\begin{document}

\title{\textbf{Energy extremals and Nonlinear Stability in a Variational theory of
a Coupled Barotropic Flow - Rotating Solid Sphere System}}
\author{Chjan C. Lim \\
Mathematical Sciences, RPI, Troy, NY 12180, USA\\
email: limc@rpi.edu\\
\\
\\
To the memory of Lim Fu Mei and Chong Yat Ying}
\maketitle

\begin{abstract}
A new variational principle - extremizing the fixed frame kinetic energy
under constant relative enstrophy - for a coupled barotropic flow - rotating
solid sphere system is introduced with the following consequences. In
particular, angular momentum is transfered between the fluid and the solid
sphere through a modelled torque mechanism. The fluid's angular momentum is
therefore not fixed but only bounded by the relative enstrophy, as is
required of any model that supports super-rotation.

The main results are: At any rate of spin $\Omega $ and relative enstrophy,
the unique global energy maximizer for fixed relative enstrophy corresponds
to solid-body super-rotation; the counter-rotating solid-body flow state is
a constrained energy minimum provided the relative enstrophy is small
enough, otherwise, it is a saddle point.

For all energy below a threshold value which depends on the relative
enstrophy and solid spin $\Omega $, the constrained energy extremals consist
of only minimizers and saddles in the form of counter-rotating states$.$
Only when the energy exceeds this threshold value can pro-rotating states
arise as global maximizers.

Unlike the standard barotropic vorticity model which conserves angular
momentum of the fluid, the counter-rotating state is rigorously shown to be
nonlinearly stable only when it is a local constrained minima. The global
constrained maximizer corresponding to super-rotation is always nonlinearly
stable. 
\end{abstract}

\newpage\ 

\section{Introduction}

The main aim of this paper is to present the simplest possible formulation
and rigorous analysis of a theory for geophysical flow coupled to an
infinitely massive rotating sphere, which exhibits characteristics of super
and sub-rotation. Although the General Circulation Model (GCM) simulations
reproduced some aspects of super-rotation such as those on Venus and Titan,
a rigorous theory which can be analysed in classical mathematical terms, is
lacking. This is the motivation which guides our approach here. We therefore
propose a mathematically precise constrained variational problem in terms of
the pseudo energy (rest frame kinetic energy) of a coupled barotropic flow
on a rotating sphere, in which the only explicit constraints are relative
enstrophy and zero total circulation. Unlike previous work, the angular
momentum of the fluid is not fixed. 

As the most promising phenomenological theory to date is one where the
rotating atmosphere exchanges angular momentum with the planetary surface
through a complex torque mechanism mediated by Hadley cells, we propose a
variational theory to find and analyze the nonlinear stability of
steady-states of the coupled fluid - sphere system. The coupled system is a
conservative or non-dissipative model in the sense that the energy and
angular momentum of the combined fluid-sphere system are separately
conserved in time. We will not go into the details of standard arguments for
the validity of an inviscid assumption for the interior flow in the context
of large scale geophysical flows. Suffice it to say at this point that the
role of viscosity is restricted to very thin Ekman boundary layers, and is
within that part of the coupled fluid-sphere system modelled by the complex
torque mechanism. 

The main results of this approach is that, at any rate of spin $\Omega $ and
relative enstrophy $Q_{rel}$, the constrained global energy maximum for
fixed relative enstrophy corresponds to solid-body rotation, $%
w_{Max}^0(Q_{rel})=\sqrt{Q_{rel}}\psi _{10},$ in the direction of $\Omega .$
Another solution, the counter-rotating steady state $w_{\min }^0(Q_{rel})=-%
\sqrt{Q_{rel}}\psi _{10},$ is a constrained global energy minimum provided
the relative enstrophy is small enough, that is, $Q_{rel}<\Omega ^2C^2$
where $C=||\cos \theta $ $||_2.$ If $\Omega ^2C^2<Q_{rel}<4\Omega ^2C^2,$
then $w_{\min }^0(Q_{rel})$ is a saddle point. And if $Q_{rel}>4\Omega
^2C^2, $ then $w_{\min }^0(Q_{rel})$ is again a saddle point, but in the
restricted sense that it is a constrained local energy maximum in all
eigen-directions except $\psi _{1,\pm 1};$ $w_{\min }^0(Q_{rel})$ is a
constrained local energy minimum in the projection onto $span(\psi _{10},$ $%
\psi _{1,\pm 1})$ where the $\psi _{1,\pm 1}$ eigenmodes are relative
vorticity patterns associated with the tilt instability.

The main result can be viewed another way. For all energy $H$ below a
threshold value $H_c$ which depends on the relative enstrophy $Q_{rel}$ and
spin $\Omega ,$ the constrained energy extremals consist of only minima and
saddles in the form of counter-rotating states, $-\sqrt{Q_{rel}}\psi _{10}.$
Only when the energy exceeds this threshold value $H_c$, can pro-rotating
states,$\sqrt{Q_{rel}}\psi _{10}$ arise as constrained global energy maxima.

It is also useful to view the approach in this paper in the light of a
vertically coarse-grained variational analysis of the steady states of the
barotropic component of fully stratified rotating flows coupled to a massive
rotating sphere. We recall another robust connection between the
time-independent variational theory presented and dynamic initial value
problems for damped driven quasi-geostrophic flows: The Principle of
Selective Decay or Minimum enstrophy states that, for suitable initial data,
the long time asymptotics of the 2D Navier-Stokes equations - damped and
forced on a doubly-periodic domain (and by extension the sphere) - is
characterized by a decreasing enstrophy to energy ratio \cite{Foias}. These
Minimum Enstrophy flows are characterized by special steady states which are
the solid body rotational flows in the case of the sphere, by virtue of an
application of Poincare's Inequality \cite{lax}. In 1953, Fjortoft \cite
{Fortoft} used a spherical harmonics expansion to show that, for dissipative
barotropic flows, because the enstrophy spectrum is related to the energy
spectrum by $Z(l,t)=l(l+1)E(l,t)$, the energy fluxes towards low wavenumbers 
$l$ while enstrophy fluxes towards high $l$ modes. His work is influential
in bringing forth the concepts of the inverse cascade of energy and forward
cascade of enstrophy that are closely related to Selective Decay.

Indeed, an inviscid asymptotic formulation of the Principle of Minimum
Enstrophy yields a variational problem that is the dual of the one in this
paper, namely, extremize enstrophy at fixed values of the total kinetic
energy of flow without fixing the angular momentum of the fluid. The
corresponding extremals including states of minimum enstrophy are again
solid-body rotating flows. Leith \cite{Leith85} formulated a related method
precisely by looking for constrained enstrophy extremals in vortex dynamics
with fixed angular momentum of the fluid.

Stability results for barotropic flows are discussed in Fjortoft \cite
{Fjortoft}, Tung \cite{Tung}, and Shepherd et al \cite{shepherdmu1}, \cite
{Wshepherd} amongst many others. Fjortoft \cite{Fjortoft} first used
integral theorems to argue that solid-body rotational flows are always
stable in the context of the Hamiltonian PDE known as the barotropic
vorticity equaton (BVE). The discrepancy between his result and those
reported here - the retrograde solid-body flow is always a stable
steady-state in the BVE but has only conditional nonlinear stability in the
coupled barotropic flow - rotating solid sphere system - is partly due to
the simple but vital fact that the fluid's angular momentum is allowed to
vary in the latter. We will elaborate on this point below.

Arnold \cite{Arnold1}, \cite{Arnold2} formalized this body of results known
as the energy-Casimir method, and proved that they work at finite
amplitudes, that is in a nonlinear way. According to an argument of Shepherd
in his refined application of the energy-Casimir method to the BVE \cite
{shepherd}, the Lyapunov function method fails precisely when the planetary
spin $\Omega $ is small - the proof of the nonlinear stability of solid-body
flow states in the BVE breaks down when the spin is small or zero. See also
the note by Wirosoetisno and Shepherd \cite{Wshepherd} which proved that
there are two important cases in 2D Euler flows where the Arnold stability
method fails, and one of them is the non-rotating sphere. Our result on the
conditional instability of the retrograde (east to west) solid-body flow and
the nonlinear stability of the prograde solid-body flow is partially
compatible with Shepherd et al because we proved instability of the
retrograde flow state only when the planetary spin $\Omega $ is smaller than 
$\sqrt{Q_{rel}/C^2},$ and because our result also states that the prograde
rigid rotation state is nonlinearly stable even when the planetary spin is
relatively small. Again the discrepancy between our results and those of
Shepherd et al can be traced to the fact that the coupled barotropic fluid
sphere model is not - unlike the BVE - a Hamiltonian PDE and angular
momentum of the fluid is not conserved.

Indeed, the correct variational theory proposed here is related to the zero
temperature version of the statistical mechanics energy - relative enstrophy
theory given in \cite{mean06}, \cite{Limsiap}. There as here, the underlying
model - a coupled fluid sphere system - cannot be represented by a
Hamiltonian or a Lagrangian. Therefore, there is no local in time equation
of motion for the underlying model but rather a generalized variational
principle operating in overall phase space determines its symmetries and
dynamical properties. This variational theory and corresponding statistical
mechanics theory comprise a new example of Feynman's generalized Least
Action Principle that is applicable to non-local situations which do not
have a Hamiltonian or Lagrangian formulation. In this setting, we propose a
non-Hamiltonian energy functional that plays a principal role in the theory
presented here and, as the integrand in the action of a partition function
(or path-integral) \cite{mean06}, in the corresponding statistical mechanics
theory as well.

This statistical mechanics theory predicts that the macrostates with
extremal internal energy - highly organized, maximally solid-body flow
states - are invariably associated with low Shannon entropy, and are
therefore the most probable states when the absolute value of the
statistical temperature is small. On the other hand, when the temperatures
have large absolute values, the most probable macrostates - disordered flow
states with little or zero net angular momentum - have intermediate values
of internal energy and high Shannon entropy \cite{mean06}. 

In section 2 we give properties of the model, as useful background for the
rest of the paper. Section 3 describes the energy-enstrophy class of
variational formulations for the coupled fluid - sphere model. Section 4
gives the constrained optimization of flow energy, $H$ on fixed relative
enstrophy manifolds and section 5 describes the physical characteristics of
the constrained energy extremals that arise in the model. Section 6 gives
the nonlinear stability results. Section 7 concerns physical consequences
and concluding remarks.

\section{Coupled Barotropic Fluid - Rotating solid Sphere Model.}

Consider the system consisting of a rotating massive rigid sphere of radius $%
R,$ enveloped by a thin shell of non-divergent barotropic fluid. The
barotropic flow is assumed to be inviscid, apart from an ability to exchange
angular momentum and kinetic energy with the infinitely massive solid sphere
through a complex torque mechanism. We also assume that the fluid is in
radiation balance and there is no net energy gain or loss from insolation.
This provides a crude model of the complex planet - atmosphere interactions,
including the enigmatic torque mechanism responsible for the phenomenon of
atmospheric super-rotation - one of the main applications motivating this
work. 

For a geophysical flow problem concerning super-rotation on a spherical
surface there is little doubt that one of the key parameters is angular
momentum of the fluid. In principle, the total angular momentum of the fluid
and solid sphere is a conserved quantity but by taking the sphere to have
infinite mass, the active part of the model is just the fluid which relaxes
by exchanging angular momentum with an infinite reservoir. It is also clear
that a 2d geophysical relaxation problem such as this one will involve
energy and enstrophy. The rest frame energy of the fluid and sphere is
conserved. Since we have assumed the mass of the solid sphere to be
infinite, we need only keep track of the kinetic energy of the barotropic
fluid - in the non-divergent case, there is no gravitational potential
energy in the fluid because it has uniform thickness and density, and its
upper surface is a rigid lid.

In a nutshell, we need to find a suitable set of constraints for the obvious
choice of objective functional, namely rest frame kinetic energy of flow in
the coupled model. The choice of auxillary conditions or constraints is not
apriori obvious, as different choices can have slightly different physical
consequences \cite{shi}.

We will use spherical coordinates - cos$\theta $ where $\theta $ is the
colatitude and longitude $\phi $. The total vorticity is given by  
\begin{equation}
q(t;cos\theta ,\phi )=\Delta \psi +2\Omega \cos \theta   \nonumber
\end{equation}
where  $2\Omega \cos \theta $ is the planetary vorticity due to spin rate $%
\Omega $ and $w=\Delta \psi $ is the relative vorticity given in terms of a
relative velocity stream function $\psi $ and $\Delta $ is the negative of
the Laplace-Beltrami operator on the unit sphere $S^2.$

\subsection{Inverse of the Laplacian}

Key to the analysis in this paper is the inverse integral operator defined
by 
\[
G[f](x)=-\frac 12\int_{S^2}dx^{\prime }f(x^{\prime })\ln \frac
1{|x-x^{\prime }|}
\]
on $S^2.$ In terms of $G,$ the solution of 
\[
\Delta \psi =f(x)
\]
is given by 
\begin{eqnarray*}
\psi (x) &=&\Delta ^{-1}(f)=G[f](x) \\
&=&-\frac 12\int_{S^2}dx^{\prime }f(x^{\prime })\ln \frac 1{|x-x^{\prime }|}.
\end{eqnarray*}

\subsection{Eigenfunctions of $G$}

The variational analysis of this problem is based on the fact that there is
an orthonormal basis for $L^2(S^2)$ of eigenfunctions of the
Laplace-Beltrami operator. And $G$ being the inverse of $\Delta ,$ its
eigenfunctions consist of the spherical harmonics, 
\begin{eqnarray}
\psi _{lm},\text{ }m &=&-l,...,l  \label{sh1} \\
\text{with eigenvalue }\lambda _{lm} &=&\text{ }-1/[l(l+1)].  \nonumber
\end{eqnarray}
Thus, a relative vorticity field, by Stokes theorem, has expansion 
\begin{equation}
w(x)=\sum_{l\geq 1,m}\alpha _{lm}\psi _{lm}(x).  \label{eexp}
\end{equation}
A key property that will be established later is that the mode $\alpha
_{10}\psi _{10}(x)$ contains all the angular momentum in the relative flow
with respect to the frame rotating at the fixed angular velocity $\Omega $
of the sphere.

\subsection{Physical quantities of the coupled barotropic vorticity model}

The rest frame kinetic energy of the fluid expressed in a frame that is
rotating at the angular velocity of the solid sphere is  
\begin{eqnarray*}
H_T[q] &=&\frac 12\int_{S^2}dx\left[ (u_r+u_p)^2+v_r^2\right]  \\
&=&\frac 12\int_{S^2}dx\left[ (u_r^2+v_r^2)+2u_ru_p\right] +\frac
12\int_{S^2}dx\text{ }u_p^2 \\
&=&-\frac 12\int_{S^2}dx\text{ }\psi q+\frac 12\int_{S^2}dx\text{ }u_p^2
\end{eqnarray*}
where $u_r,$ $v_r$ are the zonal and meridional components of the relative
velocity, $u_p$ is the zonal component of the planetary velocity (the
meridional component being zero since planetary vorticity is zonal), and $%
\psi $ is the stream function for the relative velocity.

Since the second term $\frac 12\int_{S^2}dx$ $u_p^2$ is fixed for a given
spin rate $\Omega ,$ it is convenient to work with the pseudo-energy as the
energy functional for the model, 
\begin{eqnarray}
H[w] &=&-\frac 12\int_{S^2}dx\text{ }\psi q=-\frac 12\int dx\text{ }\psi
(x)\left[ w(x)+2\Omega \cos \theta \right]   \nonumber \\
&=&-\frac 12\int dx\text{ }\psi (x)w(x)-\Omega \int dx\text{ }\psi (x)\cos
\theta   \nonumber \\
&=&-\frac 12\left\langle w,G[w]\right\rangle -\Omega C\left\langle \psi
_{10},G[w]\right\rangle   \nonumber
\end{eqnarray}
where $C=||\cos \theta ||_2.$ Later, it will be convenient to perform one
more minor adjustment of the energy functional in Lemma 3 to turn it into a
positive definite quadratic form in the Fourier coefficients of the spectral
expansion of the relative vorticity $w$ in the spherical harmonics $\psi
_{lm}.$ The following calculation using the self-adjoint and eigenvalue
properties of the integral operator $G,$ 
\begin{eqnarray*}
-C\left\langle \psi _{10},G[w]\right\rangle  &=&-C\left\langle G[\psi
_{10}],w\right\rangle  \\
&=&\frac C2\left\langle \psi _{10},w\right\rangle  \\
&=&\frac 12\int_{S^2}dx\text{ }w(x)\cos \theta (x),
\end{eqnarray*}
shows that 
\[
\Lambda [w]\equiv -C\left\langle \psi _{10},G[w]\right\rangle =\frac{\alpha
_{10}C}2
\]
is the variable net angular momentum of the flow associated with the
relative vorticity $w(x)$ (assuming unit mass density). Since 
\[
E[w]\equiv -\frac 12\left\langle w,G[w]\right\rangle =-\frac 12\int dx\text{ 
}\psi (x)w(x)
\]
is the kinetic energy of relative motion, this suggests that the
pseudo-energy $H[q]=-\frac 12\int_{S^2}dx$ $\psi q$ has the form 
\[
H[q]=E[w]+\Omega \Lambda [w]
\]
of an energy-momentum functional used in standard variational methods for
rotating problems. This is however only a coincidence.

A key difference between our approach and the standard variational analysis
of rotating problems is that neither angular momentum $\Lambda [w]$ nor
kinetic energy $H$ is fixed in our approach - the coefficient $\alpha _{10}$
in the expansion (\ref{eexp}) is not fixed. 

Total circulation in the model is fixed to be $\int q$ $dx$ $=$ $\int wdx=0,$
which is a direct consequence of Stokes theorem on a sphere and has nothing
to do with the fact that the barotropic flow is assumed to be inviscid. It
is easy to see that the kinetic energy functional $H$ is not well-defined
without the further requirement of a constraint on the size of its argument,
the relative vorticity field $w(x)$. A natural constraint for this quantity
is therefore its square norm or relative enstrophy so as to carry out the
variational analysis in the Hilbert space, $L^2(S^2).$ But before we settle
on this constraint for the variational theory, we consider the total
enstrophy of the flow, 
\[
\Gamma [q]=\int_{S^2}dx\text{ }q^2.
\]
Expanding (cf. \cite{shepherd}), 
\begin{eqnarray}
\Gamma [q] &=&\int_{S^2}dx\text{ }q^2=\int_{S^2}dx\text{ }\left[ w+2\Omega
\cos \theta \right] ^2  \nonumber \\
&=&\int_{S^2}dx\text{ }w^2+4\Omega \int_{S^2}dx\text{ }w\cos \theta +4\Omega
^2C^2,  \label{am}
\end{eqnarray}
we find that the last term is a conserved quantity because it is the square
of the $L^2$ norm of the fixed planetary vorticity, the second term is
proportional to the variable net angular momentum of the fluid (relative to
the rotating frame), and the first term is the relative enstrophy. 

From (\ref{am}), we deduce that, unlike the BVE, at most one of the two
quantities - total enstrophy and relative enstrophy - can be conserved in
the coupled fluid-sphere model. In other words, if the total enstrophy is
conserved then relative enstrophy is not because net angular momentum is not
conserved; if on the other hand, relative enstrophy is conserved, then total
enstrophy must change with net angular momentum. In fact it is not possible
- from what is assumed in the model - to exclude the third possibility that
neither the relative entrophy nor total enstrophy is conserved. 

Thus, we see that our choice of the relative enstrophy constraint, although
it is natural and required for a rigorous variational analysis, is not a
consequence of its invariance. Instead the argument that to an extent
justifies this choice is based on the Principle of Minimum Enstrophy. This
Principle states that only the ratio of enstrophy to energy - not their
separate values - should be relevant to the analysis of a quasi-steady state
in 2d flows and this ratio should have the minimum allowed value for the
given geometry. This is clearly dual and equivalent to characterizing
quasi-steady states in 2d flows in terms of their energy on iso-enstrophy
manifolds.

The second term in (\ref{am}) is equal to $4\Omega $ times the variable
angular momentum density of the relative fluid motion and has units of $m^4/s
$. The physical angular momentum, given by 
\begin{equation}
\rho \int_{S^2}dx\text{ }w\cos \theta =\rho \left\langle w,\cos \theta
\right\rangle ,  \label{am2}
\end{equation}
implies that the only mode in the eigenfunction expansion of $w$ that
contributes to its net angular momentum is $\alpha _{10}\psi _{10}$ where $%
\psi _{10}=a\cos \theta $ is the first nontrivial spherical harmonic; it has
the form of solid-body rotation vorticity.

All other moments of the vorticity $\int dx$ $q^n$ are deemed less important
in many physical theories of fluid motions including the absolute
equilibrium statistical mechanics model \cite{Kraichnan} and the variational
problems below \cite{Marsden1}. They are, however, not totally irrelevant 
\cite{shi}.

\section{Energy-relative enstrophy variational theory}

One of the aim of this approach is to answer the question: For which
relative vorticity field $w(x)$ is the fluid's rest frame kinetic energy
extremal under the single additional constraint that the square-norm of $w(x)
$ or relative enstrophy is fixed? It turns out to be physically relevant
that the $w_0(x)$ that extremizes $H[q]$ also maximizes the net angular
momentum $\Lambda [w]$ since then, these extremal steady states are super or
sub-rotating according to the sign of $\alpha _{10}$ in the expansion (\ref
{eexp}) of $w_0(x).$

A similar analysis for the non-rotating case ($\Omega =0)$ shows below that $%
H[w]=E[w]$ is maximized under fixed enstrophy by pro-rotating solid-body
flow $w_0(x)=\alpha _{10}\psi _{10}(x)$ for $\alpha _{10}>0$ which at the
same time maximizes net angular momentum $\Lambda [w_0].$ This implies
immediately the proposition:

\textbf{Proposition: }\textit{For fixed }$\Omega >0,$\textit{\ }$H[q]$%
\textit{\ is maximized under fixed relative enstrophy by the solid - body
flow }$w_0(x)$\textit{\ which also maximizes relative or net angular
momentum }$\Lambda [w_0].$

What is not obvious and therefore requires the substantial analysis
presented in this paper is the asymmetry between the energy maximizer and
mnimizer - the counter-rotating solid-body flow $w_1(x)=\alpha _{10}\psi
_{10}(x)$ for $\alpha _{10}<0$, corresponding to minimal angular momentum,
and maximizing $E[w]$ (since the energy $E[w]$ is even in $w),$ is a
minimizer of $H[q]$ for values of relative enstrophy that are small compared
to the rotation rate $\Omega ,$ but becomes a saddle point for larger values
of relative enstrophy.

Not only are the higher vorticity moments $\int dx$ $q^n$ for $n>2$ ignored
in this paper (cf. also \cite{Kraichnan}. We also do not constrain the total
enstrophy 
\[
\int_{S^2}dx\text{ }w^2+4\Omega \int_{S^2}dx\text{ }w\cos \theta 
\]
nor the angular momentum 
\begin{equation}
\int_{S^2}dx\text{ }w\cos \theta .  \label{aam}
\end{equation}
The only constraint in this paper besides the zero total circulation
condition, 
\begin{equation}
\int_{S^2}dx\text{ }w=0,  \label{tc0}
\end{equation}
is that the relative enstrophy, 
\begin{equation}
\int_{S^2}dx\text{ }w^2=Q_{rel}>0,  \label{re}
\end{equation}
is fixed.

\subsection{ Constraint $V_{rel}$}

We first analyze the physical consequences of the constraint $%
||w||_2=Q_{rel}>0$ in the definition of the subset 
\[
V_{rel}=\left\{ w\in L^2(S^2)\text{ }|\text{ }\Gamma
_{rel}[w]=||w||_2=Q_{rel}>0,\text{ }TC=\int_{S^2}dx\text{ }w=0\right\} , 
\]
which leave the angular momentum (\ref{aam}) constrained only by an
inequality. Fixing the relative enstrophy and relative circulation, this
model allows in principle energy extrema which can have up to a maximum
amount of super or sub-rotation. It has the interesting consequence that
angular momentum is effectively constrained by an inequality.\smallskip\ 

\textbf{Lemma 1}: \textit{Let the circulation and relative enstrophy of }$%
w\in L^2(S^2)$\textit{\ be constrained by (\ref{tc0}) and (\ref{re})
respectively. Then, the angular momentum (\ref{aam}) is constrained by an
inequality, that is, } 
\begin{equation}
\left| \int_{S^2}dx\text{ }w\cos \theta \right| \leq C\sqrt{Q_{rel}}
\label{ub1}
\end{equation}
\textit{where }$C>0$\textit{\ is a constant that does not depend on }$w.$%
\smallskip\ 

Proof: By the Cauchy-Schwartz inequality \cite{lax}, the result is proved
with 
\begin{equation}
C^2=\int_{S^2}(\cos \theta )^2\text{ }dx.  \label{cc}
\end{equation}
QED.\smallskip\ 

It is not surprising that solid-body rotation vorticity in the form of
spherical harmonic $\psi _{10}=b\cos \theta $ is the only zero circulation
relative vorticity field that maximizes angular momentum for given
enstrophy. This is stated precisely below:\smallskip\ 

\textbf{Lemma 2}: \textit{The upper bound on the angular momentum in (\ref
{ub1}) is achieved only for relative vorticity field } 
\[
w=k\cos \theta 
\]
\textit{where }$k$\textit{\ is a constant.}\smallskip\ 

Proof: When $w=k\cos \theta ,$ 
\[
\left| \int_{S^2}dx\text{ }w\cos \theta \right| =|k|\left| \int_{S^2}dx\text{
}\cos ^2\theta \right| =|k|C^2 
\]
and 
\[
||w||^2=\int_{S^2}dx\text{ }w^2=k^2C^2=Q_{rel}. 
\]
Thus, 
\[
\left| \int_{S^2}dx\text{ }w\cos \theta \right| =C\sqrt{Q_{rel}}. 
\]

Conversely, if 
\[
w=k\cos \theta +a\psi _{lm} 
\]
where $lm\neq 10$ and $a\neq 0,$ then 
\[
\left| \int_{S^2}dx\text{ }w\cos \theta \right| <C||w||. 
\]
QED.\smallskip\ 

\subsection{Augmented energy functionals}

The Lagrange Multiplier method for Hilbert space \cite{smith} gives
necessary conditions for constrained extremals in the form of Euler-Lagrange
equations. More importantly, the explicit form of the Lagrange Multipliers
in terms of spin, energy and relative enstrophy, provides the detailed
physical relationships that we seek in order to answer the questions raised.
We also extend the Lagrange Multiplier method to give a geometrical proof of
the existence and nonlinear stability of the constrained energy extremals\
where the constraint is fixed relative enstrophy.

The unconstrained optimization of the augmented energy-relative enstrophy
functional arising from the Lagrange Multiplier method yield necessary
conditions for extremals of the constrained optimization problems. Necessary
conditions for the extremals of the augmented functionals are in turn given
in terms of their Euler-Lagrange equations (or Gateaux derivative) which
take the form of interesting inhomogeneous linear equations involving the
inverse $G$ of the Laplace-Beltrami operator. Bifurcation values of the
multipliers can be read off the spectrum of $G$. Extremals of the augmented
energy functional typically change type for different regimes of the
Lagrange multiplier values which are marked by bifurcation values.

The relative enstrophy constraint gives rise to the following constrained
optimization problems: 
\begin{eqnarray}
&&\text{extremize }H[w]\text{ on}  \label{2a} \\
\text{ }V_{rel} &=&\left\{ w\in L^2(S^2)\text{ }|\text{ }\Gamma
_{rel}[w]=||w||_2=Q_{rel}>0,\text{ }\int_{S^2}dx\text{ }w=0\right\} .
\label{2b}
\end{eqnarray}
By the method of Lagrange Multipliers (see \cite{smith}, \cite{temam}), the
energy-relative enstrophy functional 
\begin{equation}
\text{ }E_{rel}[w;\Omega ,R]=H[w]+\lambda \Gamma _{rel}[w]  \label{2c}
\end{equation}
is the augmented objective functional for the unconstrained optimization
problem coresponding to the above constrained optimization problem.

\subsection{Lagrange Multipliers}

The Euler-Lagrange Multiplier theorem states: \textit{Let }$w_0\in V_{rel}$%
\textit{\ be an extremal of }$H[w]$\textit{. Then at least one of the
following holds:}

\[
(1)\text{ }\delta \Gamma _{rel}(w_0)=0, 
\]

\[
(2)\text{ }\delta E_{rel}(w_0)=\delta H(w_0)+\lambda \delta \Gamma
_{rel}(w_0)=0. 
\]

The right way to use this theorem is: (a) find a set of relative vorticity $%
w\in V_{rel}$ which satisfies $\delta \Gamma _{rel}(w)=0,$ (b) find a second
set of relative vorticity $w\in V_{rel}$ which satisfies $\delta
H(w)+\lambda \delta \Gamma _{rel}(w)=0,$ for some $\lambda ,$ (c) the set of
constrained extremals of $H[w]$ is contained in the union of the sets in (a)
and (b), and (d) the value of $\lambda $ is determined from the value of the
fixed constant $\Gamma _{rel}[w]=Q_{rel}>0.$

In implementing this procedure we first compute the variation or Gateaux
derivative 
\[
\text{ }\delta \Gamma _{rel}(w,\Delta w)=2\int_{S^2}w\Delta w\text{ }dx. 
\]
By choosing $\Delta w=w,$ we show that this variation never vanishes for any 
$w,$ i.e., 
\[
(1)\text{ }\delta \Gamma _{rel}(w)\neq 0. 
\]
The Euler-Lagrange Multiplier theorem then states that any extremal $w_0$ of
the constrained variational problem (\ref{2a}) and (\ref{2b}) must satisfy 
\[
(2)\text{ }\delta E_{rel}(w_0)=\delta H(w_0)+\lambda \delta \Gamma
_{rel}(w_0)=0 
\]
for some value of $\lambda .$

The vanishing of the Gateaux derivative of the augmented functional $E_{rel}$
gives the Euler-Lagrange equation for the unconstrained problem, which is
solved in the next section.

\section{Extremals of the augmented energy functional}

The analysis below will be based on the spherical harmonics (\ref{sh1})
which are eigenfunctions of the Laplace-Beltrami operator on the sphere. In
particular, the eigenfunction $\psi _{10}=a\cos \theta $ which has
eigenvalue $\lambda _{10}=-2,$ will play a special role in the
characterization of the zonal steady states. Using the linearity of $G,$ and
the eigenfunction expansion 
\begin{equation}
w=\sum_{l\geq 1,m}\alpha _{lm}\psi _{lm},  \label{exp1}
\end{equation}
of relative vorticity $w$ satisfying $\int_{S^2}dx$ $w=0,$ the Hamiltonian
functional is expanded in terms of the orthonormal spherical harmonics, to
yield

\begin{eqnarray}
H &=&-\frac 12\left\langle w,G[w]\right\rangle -\Omega C\left\langle \psi
_{10},G[w]\right\rangle  \nonumber \\
&=&-\frac 12\sum_{l\geq 1,m}\frac{\alpha _{lm}^2}{\lambda _{lm}}+\frac
12\Omega C\alpha _{10}.  \label{h2}
\end{eqnarray}
It follows that the coupling between the relative motion and the planetary
vorticity occurs through the eigenfunction $\psi _{10},$ which has the form
of a vorticity pattern corresponding to solid-body rotational flow.

For obvious technical reasons, and without loss of generality, we will use
the following lemma to change the Hamiltonian by a constant to a quadratic
form.\smallskip\ 

\textbf{Lemma 3}: \textit{For fixed spin }$\Omega >0,$\textit{\ the energy }$%
H$\textit{\ for }$w$\textit{\ satisfying } 
\[
\int_{S^2}dx\text{ }w(x)=0, 
\]
\textit{is modulo the constant } 
\begin{equation}
H_0=-\frac 14\Omega ^2C^2,  \label{ho}
\end{equation}
\textit{equal to the positive definite quadratic form (which we will again
denote by }$H)$%
\begin{equation}
H=\frac 14\left[ \alpha _{10}-(-\Omega C)\right] ^2+\frac 14\left[ \alpha
_{11}^2+\alpha _{1-1}^2\right] -\frac 12\sum_{l>1,m}\frac{\alpha _{lm}^2}{%
\lambda _{lm}}.  \label{newh}
\end{equation}
\smallskip\ 

Proof: By completing the square,

\begin{eqnarray*}
H &=&-\frac 12\sum_{l\geq 1,m}\frac{\alpha _{lm}^2}{\lambda _{lm}}+\frac
12\Omega C\alpha _{10} \\
&=&\frac 14\alpha _{10}^2+\frac 12\Omega C\alpha _{10}-\frac 12\sum_{l\geq
1,m}\frac{\alpha _{lm}^2}{\lambda _{lm}} \\
&=&\frac 14\left[ \alpha _{10}^2+2\Omega C\alpha _{10}+\Omega ^2C^2\right]
+\frac 14\left[ \alpha _{11}^2+\alpha _{1-1}^2\right] -\frac 12\sum_{l>1,m}%
\frac{\alpha _{lm}^2}{\lambda _{lm}}-\frac 14\Omega ^2C^2 \\
&=&\frac 14\left[ \alpha _{10}-(-\Omega C)\right] ^2+\frac 14\left[ \alpha
_{11}^2+\alpha _{1-1}^2\right] -\frac 12\sum_{l>1,m}\frac{\alpha _{lm}^2}{%
\lambda _{lm}}-\frac 14\Omega ^2C^2.
\end{eqnarray*}

QED.\smallskip\ 

Since all the extremals $w^0$ below have the form $k\psi _{10}$ of
solid-body rotation, it is useful to state the following result. The simple
proof is left to the reader. Graphs of (\ref{haha}) and (\ref{qrr1}) are
depicted in figures 1 and 2. \smallskip\ 

\textbf{Lemma 4}: \textit{The energy and relative enstrophy of the extremals 
}$w^0=\alpha _{10}\psi _{10}$\textit{\ takes the form} 
\begin{eqnarray}
H[\alpha _{10}\psi _{10}] &=&\frac 14\left( \alpha _{10}+\Omega C\right) ^2
\label{haha} \\
Q_{rel} &\equiv &\Gamma _{rel}[\alpha _{10}\psi _{10}]=\alpha _{10}^2. 
\nonumber
\end{eqnarray}
\textit{Furthermore, for fixed }$H,$\textit{\ } 
\begin{equation}
Q_{rel}\pm 2\Omega C\sqrt{Q_{rel}}+\Omega ^2C^2=4H  \label{qrr1}
\end{equation}
\textit{with solutions} 
\begin{equation}
Q_{rel}=\left( \pm \Omega C+\sqrt{4H}\right) ^2.  \label{qr2}
\end{equation}

\textit{\ } \smallskip\ 

\subsection{Lagrange Multipliers}

This constrained variational model is based on the fixed relative enstrophy
constraint $V_{rel}.$ By a theorem in \cite{smith} on the Lagrange
Multiplier method, the variational problem (\ref{2a}), (\ref{2b}) can be
reformulated in terms of extremals of the augmented functional, 
\begin{eqnarray*}
E_{rel}[w;\Omega ] &=&H[w]+\lambda _{rel}\Gamma _{rel}[w], \\
\Gamma _{rel}[w] &=&\int_{S^2}w^2dx.
\end{eqnarray*}
Expanding $\Gamma _{rel}[w]$ in terms of (\ref{exp1}) and combining with (%
\ref{newh}) yields 
\begin{equation}
E_{rel}[w;\Omega ]=\frac 14\left[ \alpha _{10}-(-\Omega C)\right] ^2+\frac
14\left[ \alpha _{11}^2+\alpha _{1-1}^2\right] -\frac 12\sum_{l>1,m}\frac{%
\alpha _{lm}^2}{\lambda _{lm}}+\lambda _{rel}\sum_{l\geq 1,m}\alpha _{lm}^2.
\label{eb}
\end{equation}
Taking the Gateaux derivative of $E_{rel}[w;\Omega ]$ wrt $w$ gives the
Euler-Lagrange equation, 
\begin{equation}
\lbrack G-2\lambda _{rel}](w^0)=\frac 12\Omega C\psi _{10}.  \label{el1}
\end{equation}

By the Fredholm Alternative \cite{lax}, equation (\ref{el1}) either (a) has
solutions for all values of the right hand side, or (b) has infinite number
of solutions when the right hand side is orthogonal to the kernel of $%
[G-2\lambda _{rel}].$ All the eigenvalues of $G$ are negative and form an
increasing sequence, $\lambda _{lm}^{-1}.$ There are several subcases,
namely, (1) $\lambda _{rel}$ $\in (-\infty ,$ $-\frac 14),$ (2) $\lambda
_{rel}$ $\in (-\frac 14,$ $\infty ),$ $\lambda _{rel}$\textit{\ }$\neq
-\frac 1{2l(l+1)}$ and (3) $\lambda _{rel}=-\frac 1{2l(l+1)}$\textit{\ }$\in
[-\frac 14,$\textit{\ }$0),$ which fall into the two broader classes that
depend on whether or not the kernel of $[G-2\lambda _{rel}]$ is trivial.

Although the results in the following theorem clearly have physical
significance, they are nonetheless easy to establish because the
Euler-Lagrange equation (\ref{el1}) is a linear equation. Figure 4 shows
plots of (\ref{sol1}) and (\ref{soll2}).\smallskip\ 

\textbf{Theorem 5}:\smallskip\ 

$(1)\,$\textit{Only when }$\lambda _{rel}$\textit{\ }$\in (-\infty ,$\textit{%
\ }$-\frac 14),$ \textit{can} \textit{extremal relative vorticity in the
form of solid-body rotation in the same direction as planetary vorticity
arise }$;$\smallskip\ 

(2) \textit{For }$\lambda _{rel}\in (-\frac 14,\infty ),$\textit{\ }$\lambda
_{rel}$\textit{\ }$\neq -\frac 1{2l(l+1)},$\textit{\ the extremal vorticity
(if it exists), is solid-body rotation in the opposite direction as
planetary vorticity;}\smallskip\ 

$(3)$\textit{\ Only when }$\lambda _{rel}=-\frac 1{2l(l+1)}$\textit{\ }$\in
[-\frac 14,$\textit{\ }$0),$\textit{\ can higher spherical harmonics }$\psi
_{lm},$\textit{\ }$lm\neq 10,$\textit{\ contribute to the extremal vorticity.%
}\smallskip\ 

The proof is divided into two natural subcases, namely, (a) when the kernel
of $[G-2\lambda _{rel}]$ is empty, and (b) when it is not.

\smallskip 

\textbf{(a i) }$\lambda _{rel}$\textbf{\ }$\in (-\infty ,$\textbf{\ }$-\frac
14)$

In this regime, the kernel of $[G-2\lambda _{rel}]$ is also empty, which
implies the results: 
\begin{equation}
\text{for }\lambda _{rel}\in (-\infty ,\text{ }-\frac 14),\text{ }w^0=-\frac{%
\Omega C}{2\left( \frac 12+2\lambda _{rel}\right) }\psi _{10};  \label{sol1}
\end{equation}
and, 
\begin{eqnarray*}
\text{ }w^0 &=&k\psi _{10}\text{ with }k\rightarrow 0^{+}\text{ as }\lambda
_{rel}\rightarrow -\infty ; \\
\text{ }w^0 &=&k\psi _{10}\text{ with }k\rightarrow \infty \text{ as }%
\lambda _{rel}\rightarrow -\frac 14.
\end{eqnarray*}
Thus, there is a pole-like singularity at $\lambda _{rel}=-1/4.$ This proves
part (1) of the theorem.\ 

\smallskip\ 

\textbf{(a ii) }$\lambda _{rel}$\textbf{\ }$\in ($\textbf{\ }$-\frac 14,0)$%
\textbf{\ and }$\lambda _{rel}$\textbf{\ }$\neq -\frac 1{2l(l+1)}$

In this case$,$ the following holds: 
\begin{eqnarray}
\text{for }\lambda _{rel} &\in &(-\frac 14,0),\text{ }\lambda _{rel}\text{ }%
\neq -\frac 1{2l(l+1)},  \nonumber \\
w^0 &=&-\frac{\Omega C}{2\left( \frac 12+2\lambda _{rel}\right) }\psi _{10}.
\label{soll2}
\end{eqnarray}
\smallskip\ 

(\textbf{a iii) }$\lambda _{rel}\geq 0$

Since all eigenvalues of $G$ are negative, the kernel of $[G-2\lambda
_{rel}] $ is empty for $\lambda _{rel}>0.$ Thus, the solution of (\ref{el1})
is easily found by setting $w^0=k\psi _{10}:$ 
\[
k=-\frac{\Omega C}{\left( 1+4\lambda _{rel}\right) }. 
\]
This implies the result: 
\begin{eqnarray*}
\text{ }w^0 &=&k\psi _{10}\text{ with }k\rightarrow 0^{-}\text{ as }\lambda
_r\rightarrow \infty ; \\
\text{ }w^0 &\rightarrow &-\Omega C\psi _{10}\text{ as }\lambda
_{rel}\rightarrow 0^{+}.
\end{eqnarray*}
The special solution $w^0=-\Omega C\psi _{10}$ holds when the constraint $%
V_{rel}$ is in-operative, and the variational problem is the unconstrained
optimization of the energy $H.$ Part (2) of the theorem is proved in a(ii)
and a(iii).

\smallskip\ 

(\textbf{b) }$\lambda _{rel}$\textbf{\ }$\in [-\frac 14,$\textbf{\ }$0)$%
\textbf{\ and }$\lambda _{rel}$\textbf{\ }$=-\frac 1{2l(l+1)}$

At the bifurcation values of $\lambda _{rel}$ in this regime, the following
result holds, as is easily ascertained: 
\begin{eqnarray}
\text{for }\lambda _{rel}\text{ } &=&-\frac 1{2l(l+1)}\in [-\frac 14,0),%
\text{ }l=2,3,...,  \nonumber \\
\text{ }w^0 &=&-\frac{\Omega C}{2\left( \frac 12-\frac 1{l(l+1)}\right) }%
\psi _{10}+\sum_{m=-l}^l\alpha _{lm}\psi _{lm}.  \label{sp2}
\end{eqnarray}
This proves part (3) of the theorem. QED.\smallskip\ 

\section{Existence and properties of extremals}

The final steps of the Euler-Lagrange Multiplier Method compute the value of
the multiplier $\lambda _{rel}$ by applying the fixed value of the enstrophy 
$Q_{rel}>0.$ A consequence of the completion of this circle of calculations
is the determination of the physical properties of the extremal relative
vorticity $w^0$ on the basis of the relative enstrophy $Q_{rel}$ and the
spin rate $\Omega .$ Figure 3 shows a plot of (\ref{lm1}) and (\ref{lm2}).

\textbf{Lemma 6}: \textit{The Lagrange Multipliers }$\lambda _{rel}$ \textit{%
of the extremals of the variational problem (\ref{2a}), (\ref{2b}) are given
in terms of the spin rate }$\Omega \geq 0$\textit{\ and the relative
enstrophy }$Q_{rel}>0$\textit{\ by} 
\begin{eqnarray}
\lambda _{rel}^{+} &=&-\frac 14\left[ 1+\frac{\Omega C}{\sqrt{Q_{rel}}}%
\right]  \label{lm1} \\
\lambda _{rel}^{-} &=&-\frac 14\left[ 1-\frac{\Omega C}{\sqrt{Q_{rel}}}%
\right] .  \label{lm2}
\end{eqnarray}

\smallskip

Proof: Substituting the solution (\ref{sol1}) of the Euler-Lagrange equation
into the constraint in $V_r$, 
\[
||w^0||^2=\frac{\Omega ^2C^2}{4\left( \frac 12+2\lambda _{rel}\right) ^2}%
=Q_{rel}>0, 
\]
gives the results for $\lambda _{rel}^{\pm }.$ The remaining special
solutions (\ref{sp2}) correspond to a countable set of bifurcation values.
QED.\smallskip 

The proof of the following statement is left to the reader. \smallskip

\textbf{Lemma 7}: (1) \textit{The first branch of solutions in Lemma 6, } 
\begin{equation}
w_{Max}^0=\sqrt{Q_{rel}}\psi _{10}  \label{wmax}
\end{equation}
\textit{corresponding to } 
\[
\lambda _{rel}^{+}<-\frac 14, 
\]
\textit{\ are associated with solid-body rotation in the direction of spin }$%
\Omega .$ \textit{In terms of the original kinetic energy} 
\begin{eqnarray}
H_{Max}[\alpha _{10}\psi _{10}] &=&-\Omega ^2C^2\frac{(1+8\lambda _{rel}^{+})%
}{16\left( \frac 12+2\lambda _{rel}^{+}\right) ^2}  \nonumber \\
&=&\frac 14Q_{rel}+\frac 12\Omega C\sqrt{Q_{rel}}.  \label{hmaxc}
\end{eqnarray}

(2) \textit{The second branch of solutions } \textit{\ } 
\begin{equation}
w_{\min }^0=-\sqrt{Q_{rel}}\psi _{10}  \label{wmin}
\end{equation}
\textit{associated with } 
\begin{eqnarray*}
\lambda _{rel}^{-} &\in &(-\frac 14,\infty ), \\
\lambda _{rel}^{-} &\neq &-\frac 1{2l(l+1)},l=2,3,...,
\end{eqnarray*}
\textit{correspond to} \textit{solid-body rotation opposite the direction of
spin. In this case, the original kinetic energy } 
\begin{eqnarray}
H_{\min }[\alpha _{10}\psi _{10}] &=&-\Omega ^2C^2\frac{(1+8\lambda
_{rel}^{-})}{16\left( \frac 12+2\lambda _{rel}^{-}\right) ^2}  \nonumber \\
&=&\frac 14Q_{rel}-\frac 12\Omega C\sqrt{Q_{rel}}  \label{hminc}
\end{eqnarray}
\textit{\ for given relative enstrophy }$Q_{rel}$\textit{\ and spin }$\Omega
.$

\smallskip\ 

Because the Euler-Lagrange method gives only necessary conditions, to get
sufficient conditions for the existence of constrained extremals, it is
traditional to use the so-called Direct Method of The Calculus of Variations 
\cite{smith}, \cite{temam}, \cite{lax}, where it is shown in two parts, that
(i) the unconstrained extremals of an augmented objective functional exist,
and (ii) these unconstrained extremals are the constrained extremals of the
original objective functional. This classical approach is presented in a
sequel \cite{shi}. The approach in this paper of using the infinite
dimensional geometry of the energy and relative enstrophy manifolds, can be
viewed as an alternative rigorous method for proving the existence of
constrained extremals in the coupled fluid-sphere model.

The Euler-Lagrange method gives necessary conditions for extremals of a
constrained problem. Sufficient conditions for the existence of extremals
can nonetheless be found from the geometrical basis of the method itself. An
extremal must be contained in the set of points $p$ with common tangent
spaces of the level surfaces of $H$ and $\Gamma ,$ but only those points $p$
where one level surface remains on the same side of the other level surface
in a full neighborhood of $p$ correspond to constrained energy maximizers or
minimizers. Of those points, if moreover, the level surface of $H$ is on the
outside of the surface of $\Gamma $ in a full neighborhood of $p_{Max},$
with respect to the point $p_o=(-\Omega C,0,...,0)$ in the subspace of $%
L_2(S^2)$ defined by (\ref{exp1}), then $p_{Max}$ is a constrained energy
maximum. If on the other hand, the level surface of $H$ is on the inside of
the surface of $\Gamma $ in a full neighborhood of $p_{\min },$ with respect
to the point $p_o=(-\Omega C,0,...,0),$ then $p_{\min }$ is a constrained
energy minimum.

All of the above Lemmas and Theorems form that part of variational analysis
known as Necessary Conditions. We will use these facts and lemma 3 in the
proof of the following existence result which gives Sufficient Conditions
for constrained extremals in terms of the geometry of the objective and
constraint functionals. Figure 5 depicts case (1) in Theorem 8; figure 6
depicts case (2C) and figures 7a and 7b show the retrograde solid-body state
as constrained local minimum and maximum respectively in both case (2A) and
(2B).\smallskip\ 

\textbf{Theorem 8}: \ (1) \textit{The first branch of solutions }$w_{Max}^0=%
\sqrt{Q_{rel}}\psi _{10}$\textit{\ in Lemmas 6 and 7 are global energy
maximizers\ for any relative enstrophy }$Q_{rel}$\textit{\ and spin }$\Omega
.$\textit{\ }\smallskip\ 

(2) \textit{For the second branch of solutions }$w_{\min }^0=-\sqrt{Q_{rel}}%
\psi _{10}$\textit{\ in Lemmas 6 and 7, the following statements hold: }

(A) \textit{If relative enstrophy is large compared to spin, i.e., } 
\begin{equation}
Q_{rel}>4\Omega ^2C^2  \label{highrange}
\end{equation}
\textit{then }$w_{\min }^0$\textit{\ is a special saddle point, in the sense
that, it is a local energy maxima in all eigen-directions except for }$%
span\{\psi _{1,\pm 1}\}$\textit{, where it is a local minima. }

(B) \textit{If relative enstrophy satisfies} 
\begin{equation}
\Omega ^2C^2<Q_{rel}<4\Omega ^2C^2  \label{midrange}
\end{equation}
\textit{then }$w_{\min }^0$ \textit{is a constrained energy saddle point.}

(C) \textit{If relative enstrophy is small compared to spin, i.e., } 
\begin{equation}
Q_{rel}<\Omega ^2C^2  \label{lowrange}
\end{equation}
\textit{then }$w_{\min }^0$\textit{\ is a constrained energy minima.} %
\smallskip\ 

Proof: (1) From the eigenvalues $\lambda _{lm}=$ $-l(l+1)$ and the fact that
the spherical harmonics $\psi _{lm}$ diagonalizes the energy $H,$ and Lemma
3 (cf. eqn. (\ref{newh})), we deduce that $H$ is positive-definite in $%
L_2(S^2)$, and its level surfaces are infinite dimensional unbounded
ellipsoids centered at $p_o=$ $(-\Omega C,$ $0,...0),$ 
\begin{eqnarray}
&&\text{with the shortest semi-major axes (of equal lengths)}  \label{circle}
\\
&&\text{in }span\left\{ \psi _{10},\psi _{11},\psi _{1,-1}\right\} ; 
\nonumber
\end{eqnarray}
and 
\begin{eqnarray}
&&\text{all remaining semi-major axes (associated with }  \label{circle2} \\
&&\text{azimuthal wavenumber }l\text{ greater than }1)  \nonumber \\
&&\text{have lengths }L(l)\text{ that are quadratic in }l  \nonumber \\
&&\text{and independent of the wavenumber }m.  \nonumber
\end{eqnarray}
The level surfaces of relative enstrophy are non-compact concentric spheres
centered at $0$ in $L^2(S^2),$ 
\[
||w||_2^2=\sum_{lm}\alpha _{lm}^2=Q_{rel}>0. 
\]
Together this implies that the level surface of $H$ lies on the outside (wrt 
$p_o)$ of the relative enstrophy level surface for fixed $Q_{rel}$ in a
neighborhood of the point $w_{Max}^0(Q_{rel)}=+\sqrt{Q_{rel}}\psi _{10}.$
This proves that $w_{Max}^0(Q_{rel})$ is a global constrained energy
maximizer.

(2) Using the same nomenclature in (D), we determine that when $%
Q_{rel}<\Omega ^2C^2$, at the common point $w_{\min }^0=-\sqrt{Q_{rel}}\psi
_{10},$ the relative enstrophy hypersphere lies on the outside of the energy
ellipsoid wrt $p_o;$ indeed both surfaces are locally convex with respect to
their respective centers. This implies that $w_{\min }^0$ is a constrained
energy minimum in the case (\ref{lowrange}).

When $Q_{rel}>4\Omega ^2C^2,$ it is natural to separate the common tangent
space of the ellipsoid and the sphere at $w_{\min }^0$ into two orthogonal
parts, namely, 
\begin{equation}
(a)\text{ }span\{\psi _{1,\pm 1}\}\mathit{,\ }(b)\text{ }span\text{ }\{\psi
_{lm},\text{ }l>1\}.  \label{tangent}
\end{equation}
>From the property (\ref{circle}) and (\ref{highrange}), we deduce that the
semi-major axes in $span\left\{ \psi _{10},\psi _{11},\psi _{1,-1}\right\} $
of the energy ellipsoid at $w_{\min }^0,$ have equal lengths 
\[
L=|-\sqrt{Q_{rel}}+\Omega C|>\Omega C. 
\]
The enstrophy sphere at $w_{\min }^0$ has radii of equal lengths 
\[
L_{rel}=|-\sqrt{Q_{rel}}|>2\Omega C. 
\]
From the fact that the center of this ellipsoid is at $p_o=-\Omega C\psi
_{10}$ while the sphere is at the origin, it follows that 
\[
L<L_{rel}. 
\]
This means that in (a) $span\{\psi _{1,\pm 1}\}$ of the common tangent space
(\ref{tangent}) at $w_{\min }^0$, this ellipsoid is inside the sphere wrt $%
p_o,$ which implies that $w_{\min }^0$ is a local energy minimum there. And
using (\ref{circle2}), we deduce that, in the orthogonal complement, part
(b) of this common tangent space (\ref{tangent}) at $w_{\min }^0,$ the
ellipsoid is outside the sphere wrt $p_o,$ which implies that $w_{\min }^0$
is a local energy maximum there. Thus, $w_{\min }^0$ is a special saddle
point in the case (\ref{highrange}), in the sense that, apart from $%
span\{\psi _{1,\pm 1}\}$, it is a local maximum.

When $\Omega ^2C^2<Q_{rel}<4\Omega ^2C^2,$ it follows from property (\ref
{circle}) and the unboundedness property (\ref{circle2}) of the energy
ellipsoid, that there is a critical value of the azimuthal wavenumber $%
l_{crit},$ such that in part 
\[
(a)\text{ }span\{\psi _{lm},\text{ }l\leq l_{crit}\} 
\]
of the common tangent space at $w_{\min }^0,$ the ellipsoid is inside the
sphere wrt $p_o;$ and in the orthogonal complement 
\[
(b)\text{ }span\{\psi _{lm},\text{ }l>l_{crit}\}, 
\]
the ellipsoid is outside the sphere wrt $p_o.$ Thus, $w_{\min }^0$ is a
constrained energy saddle point in the case (\ref{midrange}). QED

\section{Nonlinear Stability Analysis}

The linear stability of the steady-states $w_{Max}$ and $w_{\min }$ can be
read off the second variation of the augmented energy functional $E_{rel}.$
We are interested in proving a stronger result, namely the nonlinear or
Lyapunov stability of the energy extremals. In principle this could be done
using the geometric (convexity) properties of the relative enstrophy and
energy manifolds, recalling that the former is a sphere in $L^2(S^2)$ and
the latter is an ``ellipsoid'' in the same Hilbert space. Due to the
infinite-dimensionality of $L^2(S^2)$, and the consequent non-compactness of
the enstrophy manifold, this is a delicate undertaking.

Instead, we propose to follow the method discussed in detail in \cite
{Marsden1}, \cite{Arnold1}, \cite{Arnold2}. The first step consists of the
identification of two quadratic functionals $Q_1$ and $Q_2$ in the
perturbation $\Delta w$ such that 
\begin{eqnarray*}
Q_1(\Delta w) &\leq &H[w_{Max}+\Delta w]-H[w_{Max}]-DH_{Max}\cdot \Delta w \\
Q_2(\Delta w) &\leq &\Gamma [w_{Max}+\Delta w]-\Gamma [w_{Max}]-D\Gamma
_{Max}\cdot \Delta w
\end{eqnarray*}
Let $\left\{ \bar{\alpha}_{lm}\right\} $ denote the perturbation Fourier
coefficient based on the steady-state $w_{Max}=\sqrt{Q_{rel}}\psi _{10},$
that is, the steady state is given by 
\begin{eqnarray*}
\alpha _{10}(w_{Max}) &=&\sqrt{Q_{rel}}; \\
\alpha _{lm}(w_{Max}) &=&0\text{ for all other modes.}
\end{eqnarray*}
It is natural and convenient to choose the positive definite quadratic forms 
\begin{eqnarray*}
Q_1(\left\{ \bar{\alpha}_{lm}\right\} ) &\equiv &H[w_{Max}+\Delta
w]-H[w_{Max}]-DH_{Max}\cdot \Delta w \\
&=&\frac{\bar{\alpha}_{10}^2}4-\frac 12\sum_{lm\neq 10}\frac{\bar{\alpha}%
_{lm}^2}{\lambda _{lm}}\geq 0; \\
Q_2(\left\{ \bar{\alpha}_{lm}\right\} ) &\equiv &\Gamma [w_{Max}+\Delta
w]-\Gamma [w_{Max}]-D\Gamma _{Max}\cdot \Delta w \\
&=&\sum \bar{\alpha}_{lm}^2\geq 0.
\end{eqnarray*}
Then, $Q_1+Q_2$ is positive definite, and is clearly a norm for the
perturbation $\Delta w$ since $Q_2(\Delta w)=||\Delta w||_2^2.$ Indeed, it
is the energy-enstrophy norm of Arnold.

From the fact that at steady-state $w_{Max},$ $D(H+\Gamma )(w_{Max})\cdot
\Delta w=0,$ it follows that 
\begin{eqnarray*}
(Q_1+Q_2)(\left\{ \bar{\alpha}_{lm}(t)\right\} ) &\equiv &(H+\Gamma
)[w_{Max}+\Delta w(t)]-(H+\Gamma )[w_{Max}] \\
&=&(H+\Gamma )[w_{Max}+\Delta w(0)]-(H+\Gamma )[w_{Max}] \\
&\leq &\Delta E(0),
\end{eqnarray*}
that is, the sum $Q_1+Q_2$ is conserved along trajectories of the BVE. Thus,
we conclude that trajectories that start near $w_{Max}$ will remain nearby
for all time, that is, $w_{Max}$ has been shown to be Lyapunov stable.

The same arguments go through for the energy minimizer $w_{\min }$ when the
relative enstrophy $Q_{rel}<\Omega ^2C^2.$ And for the saddle point $w_{\min
}$ when $Q_{rel}>4\Omega ^2C^2,$ a modified argument where perturbations are
restricted to be in the orthogonal complement of $span\{\psi _{10,}$ $\psi
_{1,\pm 1}\}$\textit{\ }in\textit{\ }$L^2(S^2)$, proves that $w_{\min }$ has
some degree of nonlinear stability. This completes the proof of

\textbf{Theorem 9: }\textit{The global energy maximizer }$w_{Max}$ \textit{%
is Lyapunov stable under all conditions of spin, energy and relative
enstrophy. The constrained energy minimizer }$w_{\min }$\textit{\ is
Lyapunov stable in the low energy, low enstrophy regime, that is, when }$%
Q_{rel}<\Omega ^2C^2.$ \textit{If }$Q_{rel}>4\Omega ^2C^2$\textit{\ (high
enstrophy, high energy regime), then the saddle point }$w_{\min }$\textit{\
is nonlinearly stable to all perturbations with azimuthal wavenumber }$l\geq
2.$ \textit{In the intermediate regime, }$\Omega ^2C^2<Q_{rel}<4\Omega ^2C^2,
$ $w_{\min }$ \textit{is unstable.}

\section{Discussion and conclusions}

From the point of view of applications to planetary atmospheres, the
physical consequences of theorems 5, 8 and 9 is embodied in the following
over-arching statement:\smallskip\ 

\textit{There is a relative enstrophy threshold value }$\Omega ^2C^2$\textit{%
such that below it, the constrained energy extremals consist of the
counter-rotating local energy minimizer state }$w_{\min }^0=-\sqrt{Q_{rel}}%
\psi _{10}$\textit{\ and the pro-rotating global energy maximizer state }$%
w_{Max}^0=+\sqrt{Q_{rel}}\psi _{10}$.\textit{\ When relative enstrophy
exceeds this value,\ the pro-rotating global energy maximizer state }$%
w_{Max}^0=+\sqrt{Q_{rel}}\psi _{10}$ \textit{persists but the
counter-rotating state }$w_{\min }^0$\textit{\ becomes a constrained saddle
point. }

\textit{In other words, the counter-rotating state is a nonlinearly stable
energy extremal only for low relative enstrophy and low kinetic energy.} 
\textit{The nonlinearly stable counter-rotating steady state }$w_{\min }^0=-%
\sqrt{Q_{rel}}\psi _{10}$\textit{\ changes stability to a constrained saddle
point via a saddle-node bifurcation when the rate of spin }$\Omega $\textit{%
\ decreases past the threshold value } 
\[
\Omega _o=\frac{\sqrt{Q_{rel}}}C,
\]
\textit{with relative enstrophy fixed at }$Q_{rel}.$\textit{\ When the
relative enstropy is large enough compared to the spin rate }$\Omega >0$%
\textit{, that is, }$Q_{rel}>\Omega ^2C^2$\textit{, the only nonlinearly
stable steady state in the model is the global energy maximizer }$w_{Max}^0=%
\sqrt{Q_{rel}}\psi _{10},$ \textit{with the caveat that for }$%
Q_{rel}>4\Omega ^2C^2,$\textit{\ the counter-rotating state }$w_{\min }^0$%
\textit{\ is nonlinearly stable to zonally-symmetric perturbations.}

\textit{Thus, for any relative enstrophy and any given spin rate }$\Omega >0,
$\textit{\ a super-rotational Lyapunov stable steady state exists in the
coupled model. When }$Q_{rel}<\Omega ^2C^2,$\textit{\ both the global energy
maximizer and global energy minimizer are Lyapunov stable solid-body
rotational states.}

The range of phenomena associated with the study of atmospheric
super-rotation such as pertaining to the Venusian atmosphere (cf. \cite{gcm}
and references therein) is very complex. While we do not claim that our
theory and results explain the whole complex range of phenomena associated
with super-rotation, by going to a model of a  barotropic fluid coupled to
an infinitely massive rotating sphere - the fluid component exchanges
angular momentum as well as kinetic energy with the sphere - we have
obtained rigorous results that are directly related to atmospheric
super-rotation in both the qualitative and quantitative sense. 

We will compare our results here with those from simulations of the General
Circulation Model (GCM) as well as those obtained from intermediate level
models. The starting point of the body of work based on the GCM is a
multi-layer stratified rotating flow model. Furthermore, the system modelled
is usually a damped driven one with differential solar radiation and Ekman
damping. The best phenomenological results to date - using the GCM - are
therefore necessarily based on intensive numerical simulations.  

For the purpose of this comparison, numerical simulations of the terrestrial
GCM show that super (sub)-rotational flows behave like saturated asymptotic
steady states of a complex damped driven system (cf. \cite{gcm} and ref.
therein). The super-rotating state appears to be associated only with
sufficiently energetic flows in these simulations. Moreover, the basin of
attraction of these flows can be reasonably characterized by only a few
initial quantities such as energy and angular momentum, and the super (sub)
- rotational end states appear to be quite robust in a numerical sense. The
super-rotational state is taken to be the one where the fluid component has
positive net angular momentum - in the direction of the sphere's spin -
relative to the frame in which the sphere is fixed. The vertically averaged
barotropic component of the flow in a damped driven multi-layered
atmospheric system therefore have asymptotic quasi-steady states which are
super and sub-rotational states near to the basic zonal states $\pm $ $\psi
_{10}$ \cite{gcm}. Recent numerics (cf. \cite{gcm} and references therein)
confirm the super-rotational state is more relevant to the atmosphere of
Venus and is associated only with sufficiently high kinetic energy; the
sub-rotational state arises only for lower energies in for instance, the
atmosphere of Titan \cite{gcm}.

A more direct comparison can be made between our detailed results and those
reported by Yoden and Yamada \cite{Yoden}. In particular, they found robust
relaxation to a prograde solid-body state for small to intermediate values
of the planetary spin, and to a retrograde solid-body state when the
planetary spin is large relative to the kinetic energy in the barotropic
flow. The first is consistent with our first prediction that the prograde
solid-body state arises only when the rest frame kinetic energy of the flow
is high enough relative to the planetary spin, or equivalently, for a given
range of kinetic energy, it is allowed only for relatively small planetary
spins. The second is partially consistent with our second prediction that
the retrograde solid-body state is nonlinearly stable only for planetary
spins that are large in comparison to the relative enstrophy, if one allows
for the fact that the large anticyclonic polar vortex state is a
superposition of mainly the retrograde solid-body state and some zonally
symmetric spherical harmonics $\psi _{l0}$ with wavenumber $l<5$.

A related point vortex formulation for the inviscid dynamics of a rotating
barotropic vorticity field has been recently reported by Newton et al in 
\cite{PN}.

\begin{figure}[htbp]
\caption{ Graph of energy $H$ vs. coordinates $k=\alpha _{10}$ of extremals
as in (\ref{haha}).}\centering
\epsfig{file=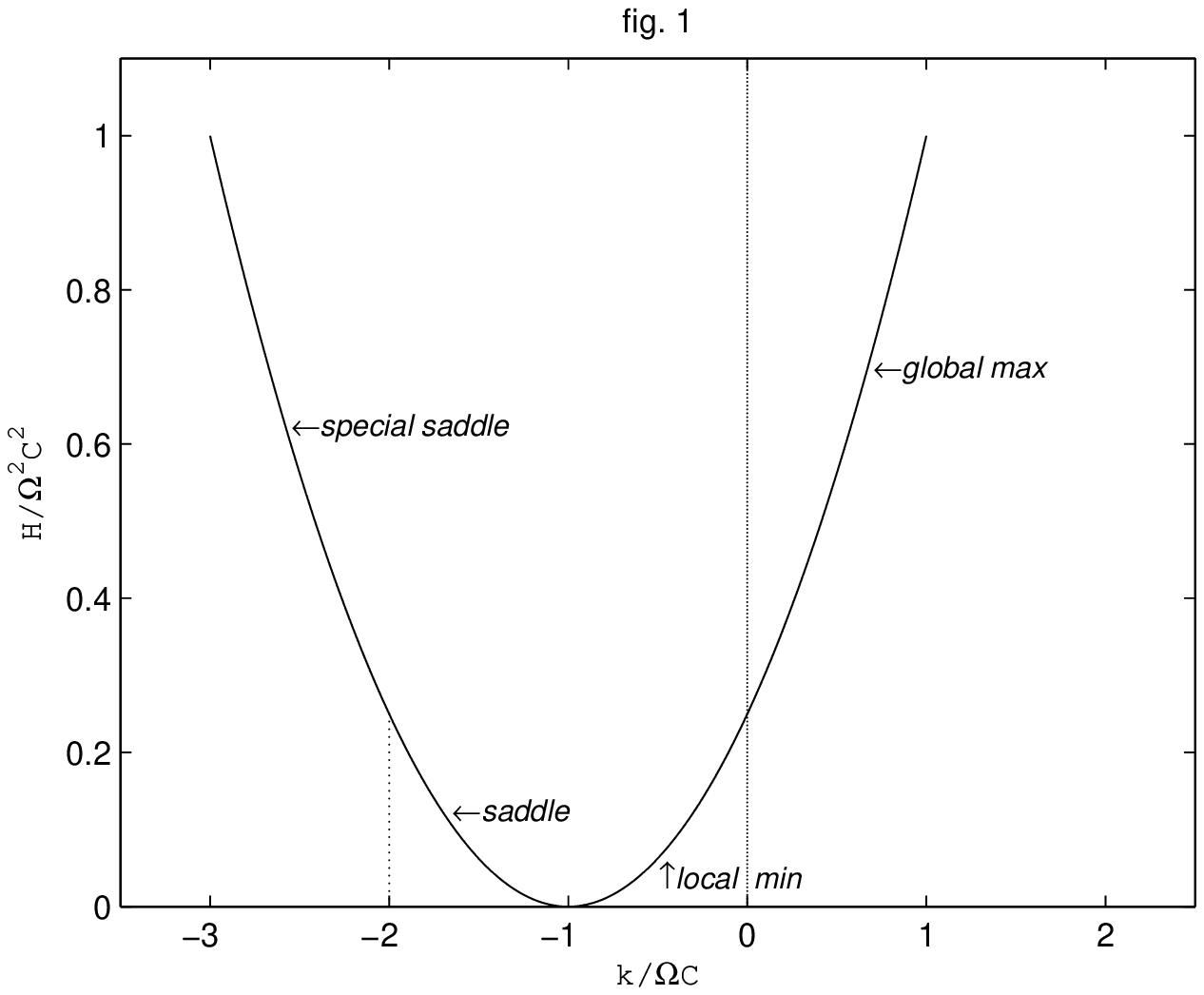,width=4.1in} \centering
\epsfig{file=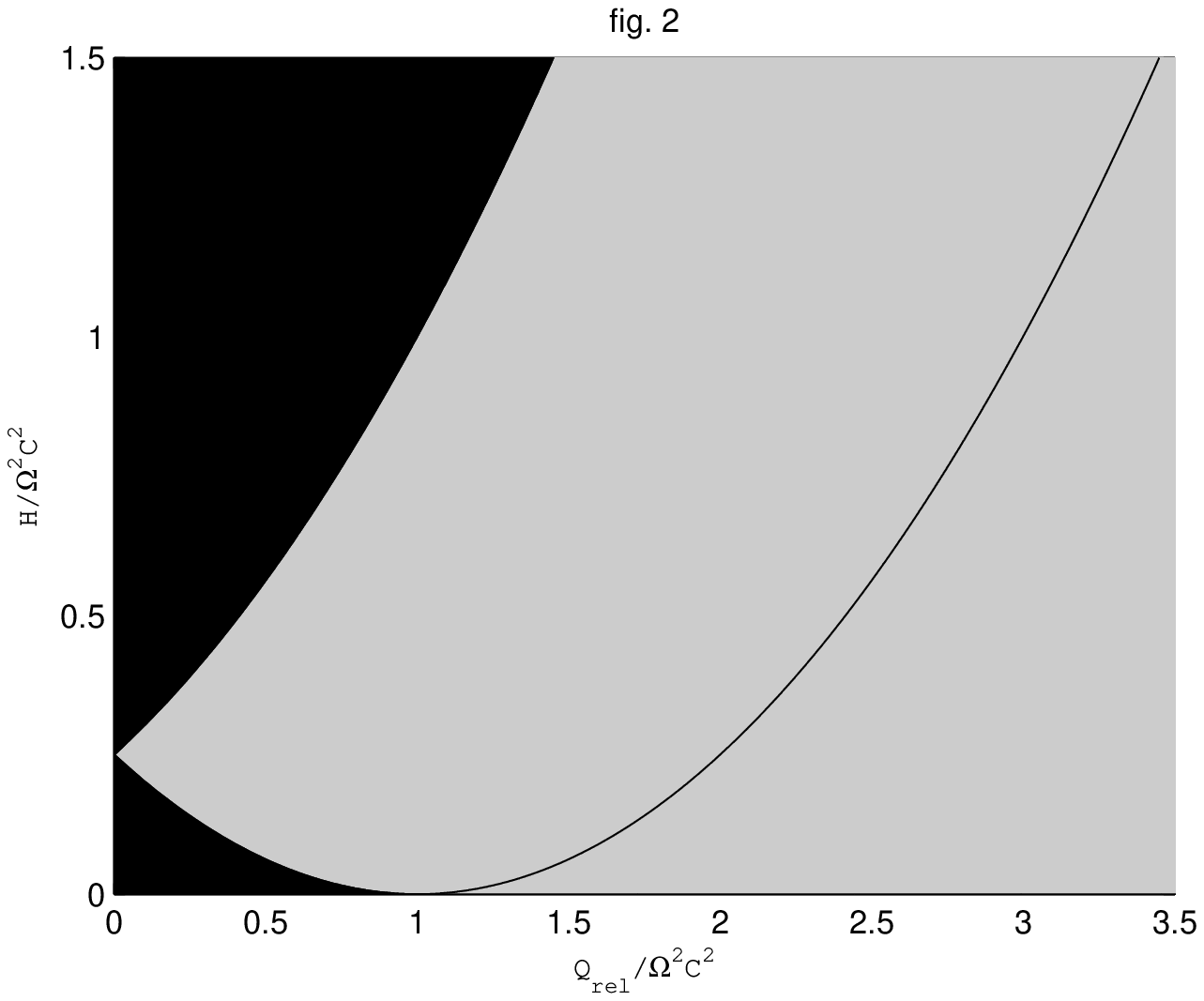,width=4.1in}
\caption{ Energy $H$ - relative enstrophy $Q_{rel}$ space; black region
denotes unpermitted values; gray region denotes non-extremal permitted
values; curves denote values at extremals as in (\ref{qrr1}).}
\end{figure}

\begin{figure}[htbp]
\caption{ Graph of Lagrange Multipliers $\lambda _{rel}^{\pm }$ vs.
square-root of relative enstrophy, $\protect\sqrt{Q_{rel}}$ for fixed spin $%
\Omega >0 $ as in (\ref{lm1}) and (\ref{lm2}).}\centering
\epsfig{file=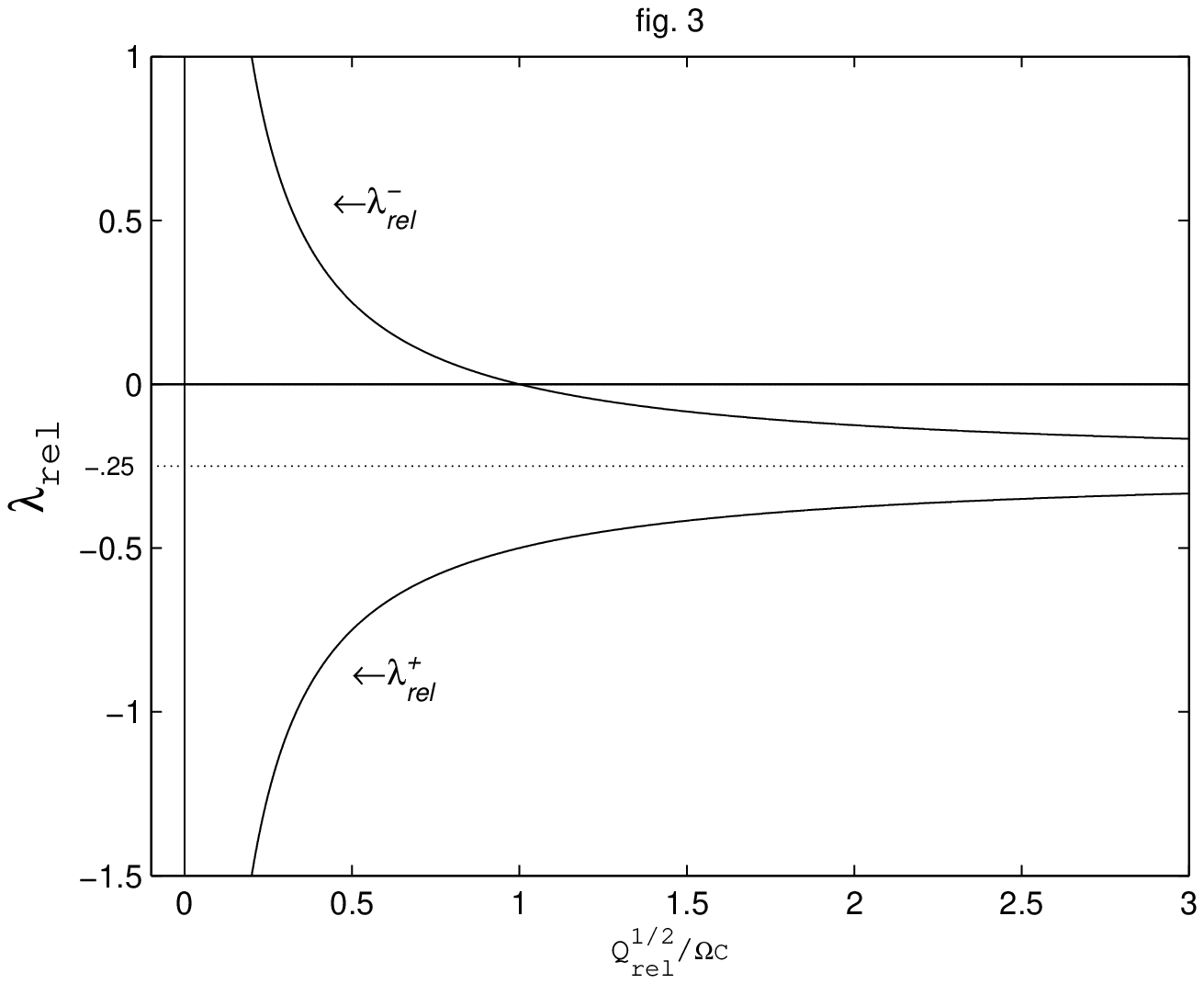,width=4.1in} \centering
\epsfig{file=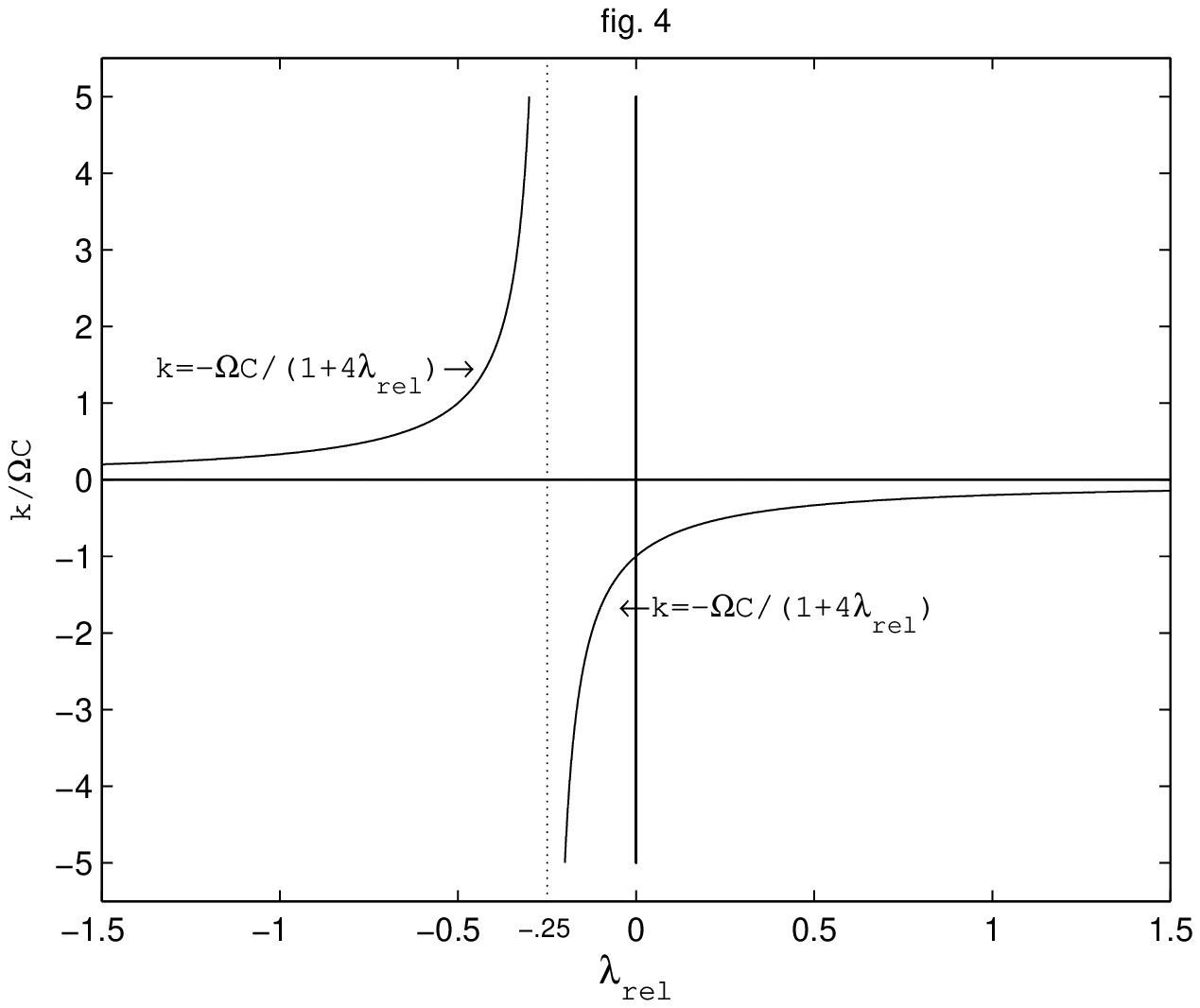,width=4.1in}
\caption{ Graph of extremal coordinates $k$ vs. Lagrange Multipliers $%
\lambda _{rel}^{\pm }$ for fixed spin $\Omega >0$ as in (\ref{sol1}) and (%
\ref{soll2}).}
\end{figure}

\begin{figure}[htbp]
\caption{ Projections of energy ellipsoid and enstrophy sphere showing the
common tangent at global maximizer $w_{Max}^0$ when energy exceeds $H_c$.}%
\centering
\epsfig{file=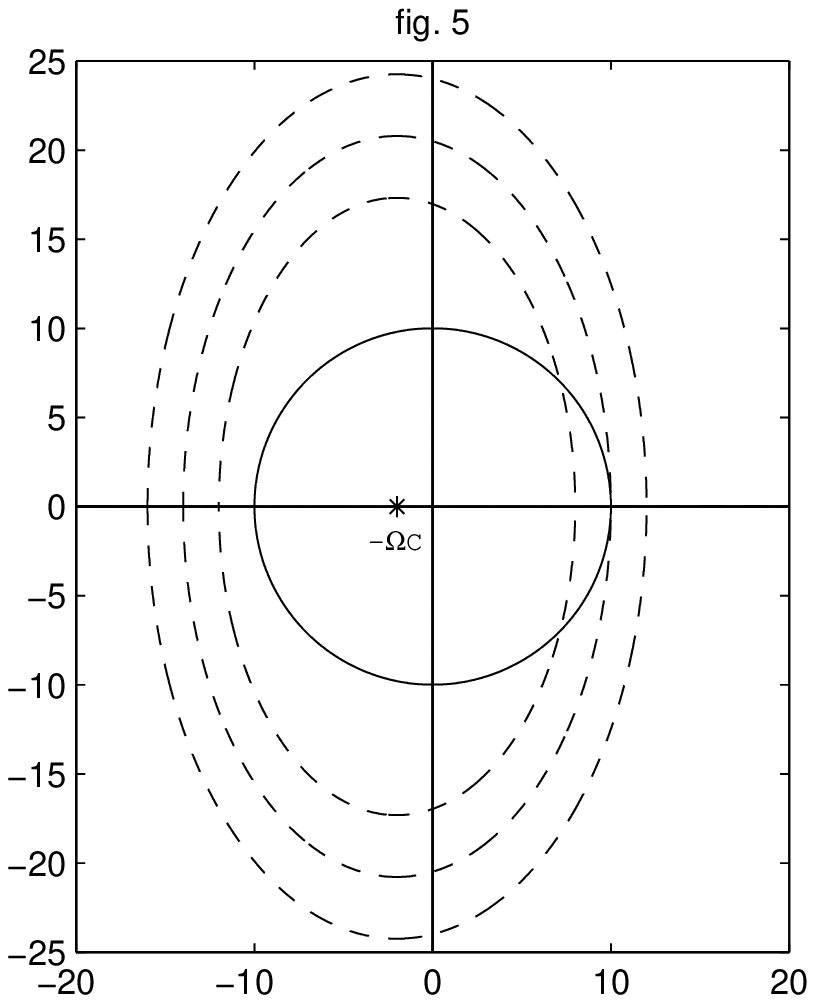,width=4.1in} \centering
\epsfig{file=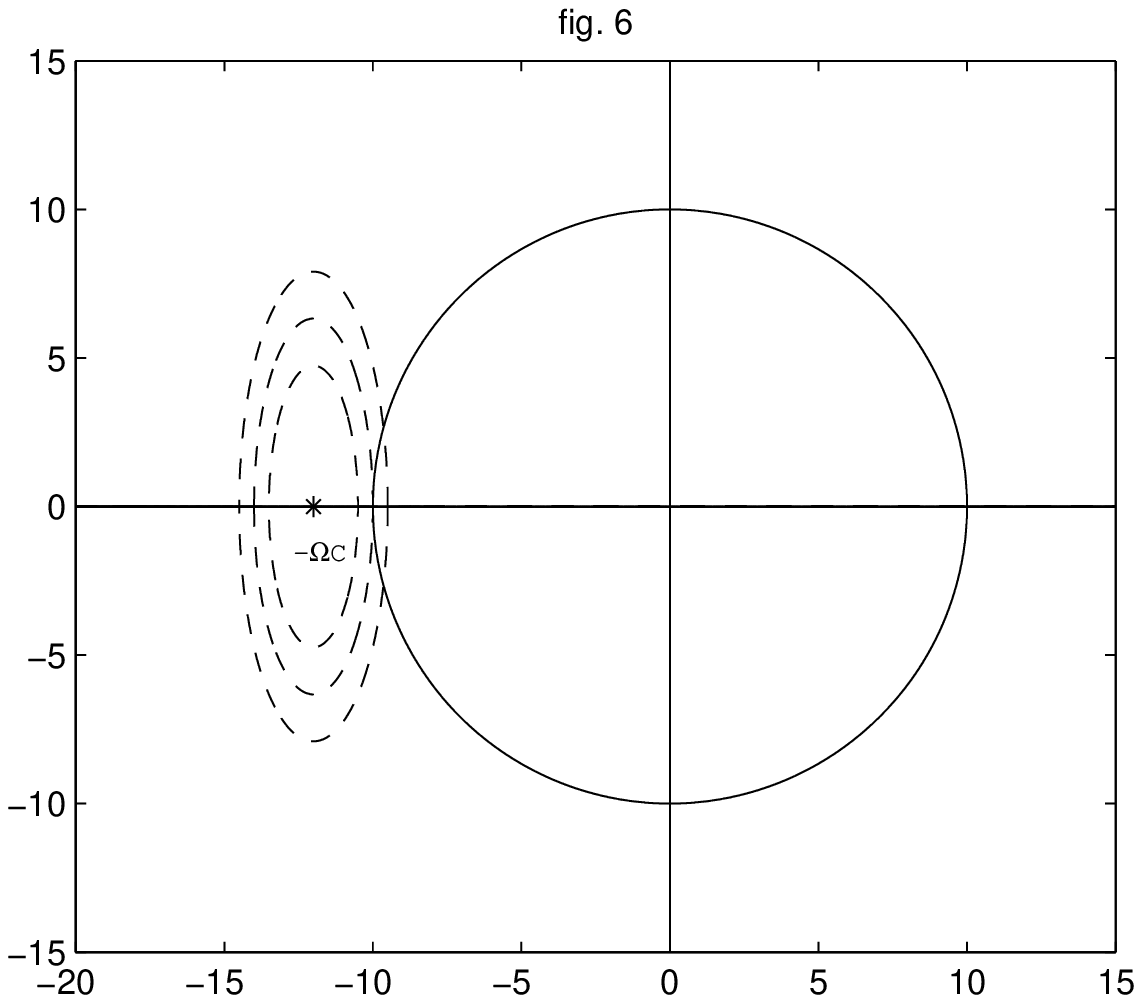,width=4.1in}
\caption{ Projections of energy ellipsoid and enstrophy sphere showing the
common tangent at local minimizer $w_{\min }^0$ when (\ref{lowrange}) holds.}
\end{figure}

\begin{figure}[htbp]
\caption{(a) Projections of energy ellipsoid and enstrophy sphere showing
the saddle point $w_{\min }^0$ as local minimum when (\ref{midrange}) or (%
\ref{highrange}) holds; (b) Projections of energy ellipsoid and enstrophy
sphere showing the saddle point $w_{\min }^0$ as local maximum }\centering
\epsfig{file=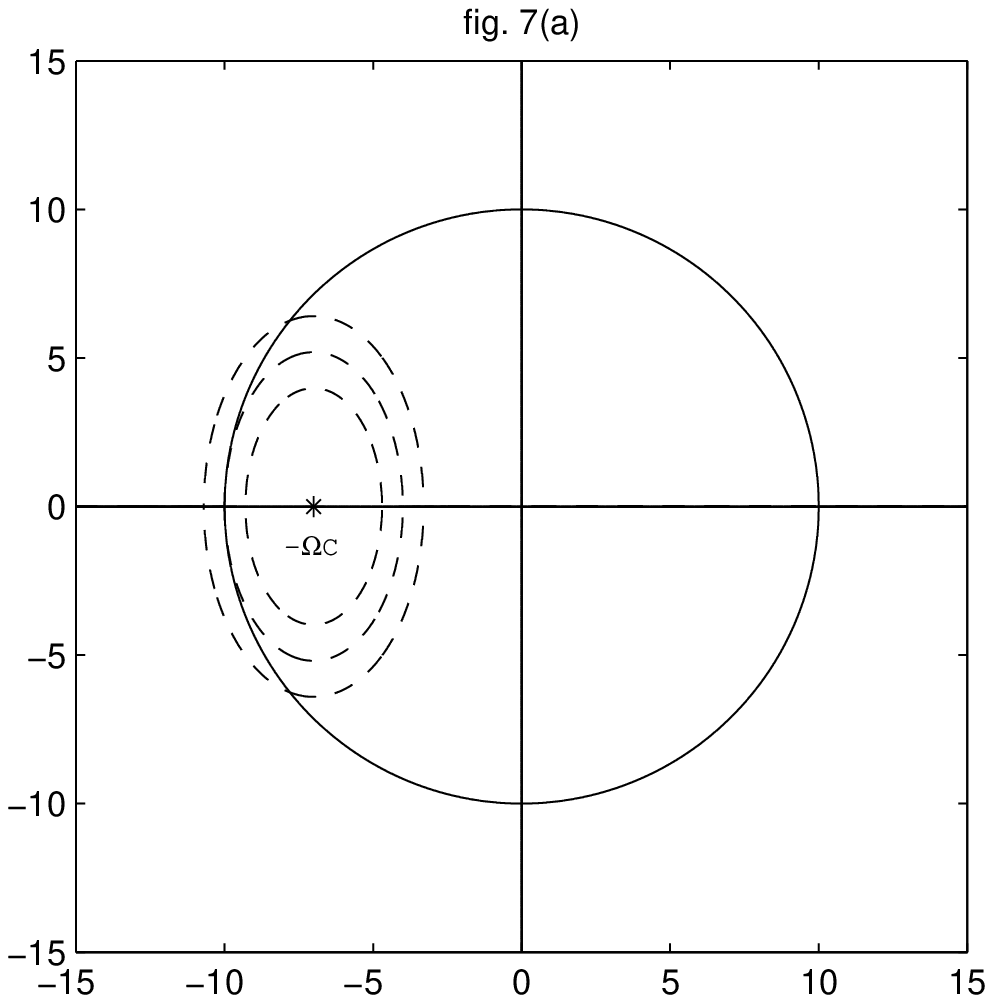,width=4.1in} \centering
\epsfig{file=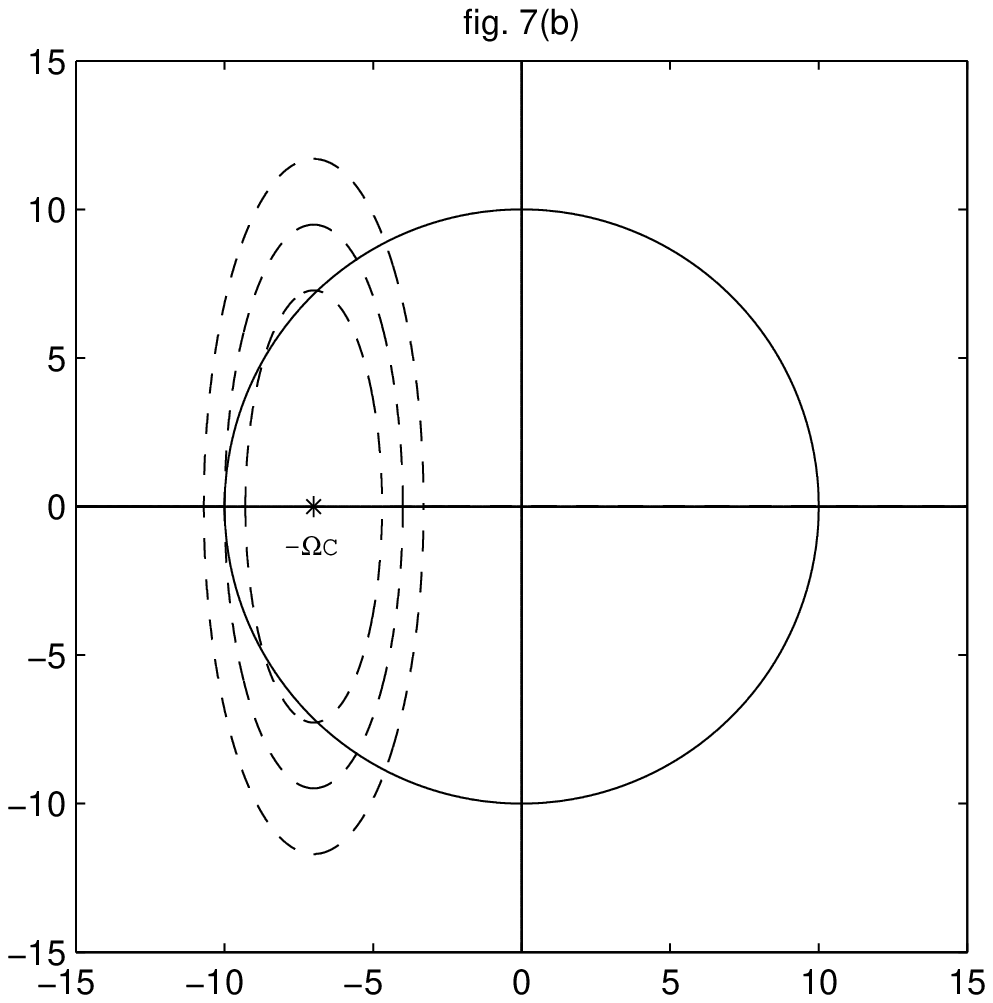,width=4.1in}
\end{figure}

Figure 1. Graph of energy $H$ vs. coordinates $k=\alpha _{10}$ of extremals
as in (\ref{haha}).

\smallskip\ 

Figure 2. Energy $H$ - relative enstrophy $Q_{rel}$ space; black region
denotes unpermitted values; gray region denotes non-extremal permitted
values; curves denote values at extremals as in (\ref{qrr1}).

\smallskip\ 

Figure 3. Graph of Lagrange Multipliers $\lambda _{rel}^{\pm }$ vs.
square-root of relative enstrophy, $Q_{rel}$ for fixed spin $\Omega >0$ as
in (\ref{lm1}) and (\ref{lm2}).

\smallskip\ 

Figure 4. Graph of extremal coordinates $k$ vs. Lagrange Multipliers $%
\lambda _{rel}^{\pm }$ for fixed spin $\Omega >0$ as in (\ref{sol1}) and (%
\ref{soll2}).

\smallskip\ 

Figure 5. Projections of energy ellipsoid and enstrophy sphere showing the
common tangent at global maximizer $w_{Max}^0$ when energy exceeds $H_c$.

\smallskip\ 

Figure 6. Projections of energy ellipsoid and enstrophy sphere showing the
common tangent at local minimizer $w_{\min }^0$ when (\ref{lowrange}) holds.

\smallskip\ 

Figure 7a. Projections of energy ellipsoid and enstrophy sphere showing the
saddle point $w_{\min }^0$ as local minimum when (\ref{midrange}) or (\ref
{highrange}) holds.

\smallskip\ 

Figure 7b. Projections of energy ellipsoid and enstrophy sphere showing the
saddle point $w_{\min }^0$ as local maximum.


\newpage

\begin{thebibliography}{99}
\bibitem{shepherd}  T. G. Shepherd, \textit{Non-ergodicity of inviscid
two-dimensional flow on a beta-plane and on the surface of a rotating sphere}%
, J. Fluid Mech., 184, 289--302, 1987.

\bibitem{mean06}  C.C. Lim, \textit{\ Extremal free energy in a simple mean
field theory for a Coupled Barotropic Fluid - rotating solid sphere system, }%
to appear DCDS - A 2007.

\bibitem{shepherdmu1}  Mu Mu and T.G. Shepherd, \textit{On Arnol'd's second
nonlinear stability theorem for two-dimensional quasi-geostrophic flow, }%
Geophys.Astrophys.Fluid Dyn. 75 21-37.

\bibitem{Cho}  J. Cho and L. Polvani, \textit{The emergence of jets and
vortices in freely evolving, shallow-water turbulence on a sphere, }Phys
Fluids 8(6), 1531 - 1552, 1995.

\bibitem{Yoden}  S. Yoden and M. Yamada, \textit{A numerical experiment on
2D decaying turbulence on a rotating sphere, }J. Atmos. Sci., 50, 631, 1993

\bibitem{Foias}  C. Foias and R. Saut, Indiana\ U. Math Journ 33, 457 - ,
1984.

\bibitem{Fjortoft}  R. Fjortoft, \textit{Application of integral theorems in
deriving criteria of stability for laminar flows and for the baroclinic
circular vortex, }Geofys. Publ. 17(6), 1, 1950.

\bibitem{Fortoft}  R. Fjortoft, \textit{On the changes in the spectral
distribution of kinetic energy for 2D non-divergent flows, }Tellus, 5, 225 -
230, 1953.

\bibitem{Leith85}  C. Leith, \textit{Minimum enstrophy vortices, }Phys.
Fluids, 27, 1388 - 1395, 1984.

\bibitem{Wshepherd}  D. Wirosoetisno and T.G. Shepherd, \textit{On the
existence of 2D Euler flows satisfying energy-Casimir criteria, }Phys.Fluids
12, 727, 2000.

\bibitem{Arnold1}  V.I. Arnold, \textit{Conditions for nonlinear stability
of stationary plane curvilinear flows of an ideal fluid, }Sov. Math Dokl.,
6, 773-777, 1965.

\bibitem{Arnold2}  V.I. Arnold, \textit{On an a priori estimate in the
theory of hydrodynamical stability, }AMS Translations, Ser. 2, 79, 267-269,
1969.

\bibitem{gcm}  A. Del Genio, W. Zhuo and T. Eichler, \textit{Equatorial
Superrotation in a slowly rotating GCM: implications for Titan and Venus, }%
Icarus 101, 1-17, 1993.

\bibitem{Chern}  S.J.Chern, \textit{Math Theory of the Barotropic Model in
GFD, }PhD thesis, Cornell University, 1991.

\bibitem{Kraichnan}  R.H. Kraichnan, \textit{Statistical dynamics of
two-dimensional flows,} J. Fluid Mech. 67, 155-175, 1975.

\bibitem{Limsiap}  C. C. Lim, \textit{Energy maximizers and robust symmetry
breaking in vortex dynamics on a nn-rotating sphere,} SIAM J. Appl Math,
65(6), 2093-2106, 2005.

\bibitem{Marsden1}  D. Holm, J. Marsden, T. Ratiu and A. Weinstein, \textit{%
Nonlinear stability of fluid and plasma equilibria, }Physics Report 123,
1-116, 1985.

\bibitem{PN}  P. Newton and H. Shokraneh, \textit{The N-Vortex Prob on a
Rotating Sphere}, Proc. R. Soc. London Series A, 462, 149 - 169, 2006

\bibitem{Rhines}  P.B. Rhines, \textit{Waves and turbulence on a beta plane, 
}J. Fluid Mech., 69(3), 417-443, 1975.

\bibitem{temam}  I. Ekeland and R. Temam, \textit{Convex Analysis and
Variatonal Problems, }North-Holland, 1976.

\bibitem{smith}  D. Smith, \textit{Variational Methods in Optimization, }%
Dover

\bibitem{Tung}  K.K. Tung, \textit{Barotropic instability of zonal flows, }%
J. Atmos. sc., 38, 308 - 321, 1981.

\bibitem{lax}  P. D. Lax, \textit{Functional Analysis}, Wiley-Interscience,
2002.

\bibitem{shi}  C.C. Lim and Junping Shi, \textit{The role of higher
vorticity moments in a variational formulation of a Barotropic Vorticity
Model on the rotating sphere}, submitted 2005.

\textbf{List of Figure Captions:}
\end{thebibliography}
\end{document}